# Evidence for coeval Late Triassic terrestrial impacts from the Rochechouart (France) meteorite crater

Laurent Carporzen[*] and Stuart A. Gilder

Institut de Physique du Globe de Paris, Equipe de Paléomagnétisme, 4 place Jussieu, 75252 Paris Cedex 05 France   [*]Corresponding author: lcarpo@ipgp.jussieu.fr

**Abstract**

High temperature impact melt breccias from the Rochechouart (France) meteorite crater record a magnetization component with antipodal, normal and reverse polarities. The corresponding paleomagnetic pole for this component lies between the 220 Ma and 210 Ma reference poles on the Eurasian apparent polar wander path, consistent with the 214±8 Ma $^{40}Ar/^{39}Ar$ age of the crater. Late Triassic tectonic reconstructions of the Eurasian and North American plates place this pole within 95% confidence limits of the paleomagnetic pole from the Manicouagan (Canada) meteorite impact crater, which is dated at 214±1 Ma. Together, these observations reinforce the hypothesis of a Late Triassic, multiple meteorite impact event on Earth.

## 1. Introduction

The speed and energy associated with meteorite impacts make them exceptional events on Earth. Although their potential to wipe out large populations of species is widely reported, details surrounding the cratering process are more ambiguous. One hotly debated topic surrounds five Late Triassic, terrestrial craters whose age distribution led *Spray et al.* [1998] to propose that they formed due to the fragmentation of a single meteorite, like that observed on Jupiter in 1994 [*Orton et al.*, 1995]. Evidence supporting the *Spray et al.* [1998] hypothesis for a fragmented, or Shoemaker-Levy 9-type, impact relied mostly on radiometric dating of the Manicouagan (Canada) and Rochechouart (France) craters at 214±1 Ma (U-Pb [*Hodych and Dunning*, 1992]) and 214±8 Ma ($^{40}$Ar/$^{39}$Ar [*Kelley and Spray*, 1997]), respectively. Late Triassic ages for the other three craters were loosely constrained by biostratigraphy. Controversy arose when *Kent* [1998] pointed out that the Manicouagan melt rocks recorded only normal magnetic polarities [*Larochelle and Currie*, 1967; *Robertson*, 1967], while those of Rochechouart had only reverse polarities [*Pohl and Soffel*, 1971]. The polarity differences thus put into question whether the two craters were truly coeval, as *Kent* [1998] reasoned that at least a few thousand years, the time needed for the Earth's field to reverse, separated the two impacts. *Spray* [1998] countered that differences in thermal mass between the two craters, owing to the much larger size of Manicouagan (100 km in diameter with a >230 m-thick melt sheet) than Rochechouart (25 km in diameter with a 12 m-thick impact melt breccia), would lead to different cooling rates, thus offsetting the time when the magnetic minerals in the rocks at the two craters cooled through their Curie points. Thus, the impact could have occurred during a reversed magnetic polarity that was recorded by the relatively fast-cooling suevites at Rochechouart; by the time the rocks cooled below ~600°C at Manicouagan, the Earth's field was in a normal polarity state.

In order to further test the Late Triassic, multiple impact event hypothesis, we re-investigated the paleomagnetism of the Rochechouart crater (Fig. 1a). We filled gaps in the previous study of *Pohl and Soffel* [1971] by sampling basement rocks and impact breccias (suevites) having experienced the highest degrees of melting. We find that the Rochechouart suevites record dual magnetic polarities and that the virtual paleomagnetic poles (VGPs) from Rochechouart are indistinguishable with those from Manicouagan in a Late Triassic reference frame.

## 2. Paleomagnetic results

We sampled 52 cores at 12 sites using a gas-powered drill. The cores were oriented with magnetic and sun compasses, the latter to correct for local declination anomalies. Four sites were drilled in impact melt breccias, while eight sites come from basement rocks of variable lithologies lying within a radius of 16 km from the crater's estimated center at 45.825°N, 0.785°E, which lies near the town of Rochechouart [*Lambert*, 1977a and 1977b] (Fig. 1a; Table 1). Samples cut from the cores underwent stepwise alternating field (AF) demagnetization up to 150 mT or thermal demagnetization up to 700°C with an average of 14 demagnetization steps. Magnetization components and mean directions were determined using principal component analysis [*Kirschvink*, 1980] and *Fisher* [1953] statistics, respectively.

After the first few demagnetization steps, the impact breccias with the highest degrees of melting, or so-called Babaudus-type suevites from sites 6 and 8, reveal stable demagnetization trajectories that decay univectorially toward the origin on orthogonal demagnetization plots (Figs. 1b to 1d). AF demagnetization removes only ~50% of the original remanence with total unblocking occurring at temperatures of 680°C, indicative of hematite as the carrier of the magnetic remanence. Thermomagnetic experiments on the same rocks reveal a Hopkinson peak and an inflection at 580°C, signaling the additional presence of single domain magnetite. All samples from Site 8 have reversed polarity magnetizations (Fig. 1b), whereas those from Site 6 possess normal polarity magnetization directions (Fig. 1c). These normal polarity samples have demagnetization characteristics identical to samples with reverse polarities, and we find no evidence (antipodal magnetization components in a single sample, etc.) to support a self-reversal mechanism to explain the polarity differences. Thermal demagnetization of two of the five samples from the Babaudus-type suevites at Site 7, located ca. 50 m from Site 6, have a component with easterly declinations and shallow inclinations that decays linearly toward the origin. Three samples subjected to AF demagnetization have two magnetic components: one resembling the two others from Site 7 at fields below 22 mT, and another isolated above 22 mT, with the same normal polarity direction observed in Site 6 (Fig. 1d). One difference between sites 6 and 7 is that the natural remanent magnetization (NRM)/bulk susceptibility is twice as low for the former (171±43 $Am^{-1}X^{-1}$, N=5) than for the latter (379±48 $Am^{-1}X^{-1}$, N=4). These values, combined with the fact that AF demagnetization more efficiently removes isothermal remanent magnetizations [*Coe et al.*, 2004], suggest that Site 7 was struck by lightning. Thus, we combined the high coercivity component of the three samples from Site 7 with those from Site 6 to calculate an overall mean direction (Table 1).

Impact breccias at the Montoume quarry (Site 5) contain decameter-size clasts of basement material (gneiss), so we drilled six samples in the melted matrix and six in the clasts in order to apply a conglomerate test. The demagnetization characteristics and magnetic mineralogies of the matrix and clasts are similar to one another and resemble those of the Babaudus-type suevites (Figs. 1e and 1f). Magnetization directions from the matrix and clasts from Montoume are indistinguishable at 95% confidence limits (negative conglomerate test [*McFadden and Lowes*, 1981]), suggesting that the clasts were heated above the hematite Curie temperature (680°C).

The magnetization directions of the melt rocks are well clustered both within sites and among the various sites (Fig. 2a), with the normal and reversed polarity directions being statistically antipodal within 95% confidence limits (Class B [*McFadden and McElhinny*, 1990]). We interpret these directions as recording a magnetic field reversal during cooling (e.g., a primary, thermal remanent magnetization). However, *Lambert* [1974 and 1977b] mapped the Rochechouart impact melt sheet as covering an area of 80 km² with a medium thickness of 12 m. He estimated that the original thickness was 100 m over 280 km², which corresponds to a characteristic cooling time of ~400 years [*Onorato et al.*, 1978]. Although this seems too rapid to have recorded a field reversal (reversals are thought to last at least a few thousand years, e.g., *Quidelleur et al.* [2003]; *Clement* [2004]), the thickness and cooling time are likely underestimated.

The basement rocks display diverse demagnetization behavior. At the Champagnac quarry (Sites 1 and 2), granodiorite, leucogranite and pseudotachylite were drilled in two places separated by a few hundred meters. The basement rocks at Champagnac lie ~10 m below the impact breccias and contain ample evidence of shock, including the presence of planar deformation features [*Kraut and French*, 1971; *Lambert*, 1974 and 1977b; *Reimold et al.*, 1984; *Bishoff and Oskierski*, 1987]. Demagnetization of one sample yielded highly erratic directions. After removing a recent field direction, magnetization directions of the sole pseudotachylite sample are noisy but mostly north and upward directed above 10 mT. Except for the pseudotachylite sample and the one incoherent sample, the rest display relatively coherent magnetization directions with southwest-oriented declinations and shallow-upward inclinations that are unblocked by 350°C (Fig. 1g). The directions isolated from sites 1 and 2 are comparable, so we combined them when calculating a site-mean direction (Table 1, Fig. 2a). A rock magnetic investigation of the Site 1&2 samples identifies a magnetic mineral characteristic of magnetite with a Hopkinson peak and a Curie temperature around 570°C, yet no evidence for a phase with a Curie temperature around 350°C, such as pyrrhotite. Moreover, hysteresis loops measured at 200°C before and after heating to 400°C [*Henry et al.*, 2005] are identical, confirming the absence of a magnetic phase with a Curie point between 200°C and 400°C. Our preferred explanation is that the Champagnac basement was heated to 350°C during impact, which remagnetized the magnetite grains capable of acquiring remanences up to 350°C. This would explain why nearly all of the remanence unblocks in a very narrow temperature range around 350°C, despite the absence of a magnetic mineral with this characteristic Curie point. If true, this would constrain the maximum temperature experienced by these basement rocks. Although the Site 1&2 mean direction has shallower inclinations than the Babaudus-type suevites, they lie close to the direction isolated in other suevite types as will be discussed below.

Gneiss, schist and microgranite samples from the Champonger quarry (Site 3) vary markedly from sample to sample. Some have erratic directions from one demagnetization step to the next. Others yield more coherent magnetization directions that fall near the expected axial dipole field direction (inclination = 64°), which lies within a few degrees of the present-day Earth's field direction. A few have noisy remanences that are often southwest and up directed but without linear trajectories that can be interpreted with confidence. Serpentinites from the Merlis quarry (Site 4) possess stable magnetizations, carried mostly by Fe-pure magnetite (Curie temperature of 580°C), with well-grouped magnetization directions (Figs. 1h and 2a, Table 1). Gneiss sampled near Fonceverane (Site 9), located a few 100 meters from the Site 8 suevites, has unstable magnetizations. Thermal and AF demagnetization of the granite collected along road D160 (Site 10) isolate north and down magnetization directions similar to the present Earth's magnetic field direction, which we interpret as a recent remagnetization. Granite from the Chabanais quarry (Site 11) has very stable magnetizations carried mainly by magnetite whose directions are reasonably well clustered (Figs. 1i and 2a, Table 1). Finally, granite sampled along Highway N141 (Site 12), northeast of the crater, possesses unstable magnetizations with weak intensities ($<5 \times 10^{-5}$ A/m).

## 3. Interpretation and Discussion

For the impact breccias at Rochechouart, the overall mean direction based on 25 samples (20 cores) is declination (D)= 36.0°, inclination (I)= 43.0°, the best estimate of the precision parameter (k)= 107.3, and the radius that the mean direction lies within 95% confidence ($\alpha_{95}$)= 2.8°. The overall mean, based on the three site-mean directions (sites 5, 6&7 and 8), is D= 35.6°, I= 42.7° (k= 806.9, $\alpha_{95}$= 4.3°), whose pole lies at latitude ($\lambda$)= 54.6°N, longitude ($\phi$)= 114.9°E ($A_{95}$= 5.2°). The previous paleomagnetic study by *Pohl and Soffel* [1971] at Rochechouart calculated a mean direction based on 130 samples (33 cores) of D= 46.4°, I= 34.8° (k= 310, $\alpha_{95}$= 4°), which is different at 95% confidence limits from the direction we obtained. Figure 2b shows the virtual geomagnetic poles (VGPs) of their sites and ours, where it appears that the VGPs lie along a swath, with the sites comprising Babaudus-type suevites lying close to the Late Triassic paleomagnetic poles from the Eurasian apparent polar wander path (APWP).

What can account for the differences between the impact breccia VGPs? Sampling techniques were virtually the same in both studies, and although *Pohl and Soffel* [1971] applied a blanket correction of -6° to account for the local declination anomaly, which is perhaps a few degrees too westerly for the late 1960s at Rochechouart [*Alexandrescu et al.*, 1999], subtracting a few degrees cannot account for the discrepancy between their study and ours. The overall mean of *Pohl and Soffel* [1971] was based only on NRM directions and not on principal component analyses derived from stepwise demagnetization. However, *Pohl and Soffel* [1971] performed magnetic viscosity tests and they compared the mean direction based on blanket AF demagnetization with the NRM directions, finding no compelling reason to reject the NRM directions. We confirm their analysis in that the mean direction based on the NRMs from our study (D= 35.4°, I= 43.9°, N= 22, k= 46.7, $\alpha_{95}$= 4.6°, excluding Site 7) is virtually identical to the one based on principal component analysis.

So either a systematic bias exists between the two laboratories or another explanation must be sought. One possibility could lie in secular variation. Indeed, three types of impact breccias, or so-called suevites, have been identified at Rochechouart based on their degrees of melting, going from Chassenon to Montoume to Babaudus types with increasing degrees of melting, respectively [*Lambert*, 1974 and 1977b; *Chèvremont et al.*, 1996]. *Pohl and Soffel* [1971] primarily sampled the Chassenon to Montoume type suevites, while we collected Montoume and Babaudus types. If the degree of melting is related to cooling time, then it is possible that the Rochechouart VGPs track secular variation of the Late Triassic geomagnetic field. The suevites that cooled the slowest would probably average secular variation, which would explain why the Babaudus type suevites recorded a reversal. It would also explain why they lie closest to the poles defining the Eurasian apparent polar wander path (APWP), which is thought to represent the time-averaged field. Of note is that the VGPs of sites 5 to 8 lie between the 210 and 220 Ma reference poles (Figure 2b), which coincides well with the $^{40}Ar/^{39}Ar$ age of 214±8 Ma for the impact [*Kelley and Spray*, 1997]. Moreover, if one corrected the 212 Ma St. Audrie's Bay (United Kingdom) Late Triassic pole for inclination shallowing [*Kent and*

*Tauxe*, 2005], the corrected pole lies within 95% confidence limits of our Rochechouart pole. Thus, we think the differences between the virtual paleomagnetic poles of *Pohl and Soffel* [1971] and our own is that the former recorded instantaneous fields whereas the latter are representative of a time-averaged field.

Magnetic data from the basement rocks may shed light on the thermal effects of the impact at Rochechouart. That all sites yielding stable magnetic remanences have reversed polarities could suggest that they were thermally remagnetized during the impact. Alternatively, because the basement rocks have radiometric ages ranging from ca. 360 to 240 Ma [*Lambert*, 1974; *Reimold and Oskierski*, 1987a and 1987b; *Chèvremont et al.*, 1996], some lie within the Kiaman reversed polarity superchron that lasted from ~315 to 260 Ma [*Opdyke and Channell*, 1996]. Figure 2b shows a polar projection of the Rochechouart VGPs from all lithologies. VGPs of Sites 1&2 and 11 lie on the far side of the VGP swath defined by the suevites and, as discussed above, the closest VGP to the suevites (Site 1&2) is particularly noteworthy because it comes from the Champagnac quarry where pseudotachylites and planar deformation features in quartz have been described [*Kraut and French*, 1971; *Lambert*, 1974 and 1977b; *Reimold et al.*, 1984; *Bishoff and Oskierski*, 1987]. Here, there is evidence that these rocks were heated to 350°C (Fig. 1g). Using a similar logic, one could suggest that the granites from the Chabanais quarry (Site 11) were heated above magnetite Curie temperatures. However, assuming that the Chabanais granite has not been tilted, its magnetic remanence is likely primary (pre-impact) because the paleolatitude coincides with that expected from the ca. 260 Ma Eurasian APWP [*Van der Voo*, 1993; *Torsvik et al.*, 2001], thus potentially being Kiaman in age. Further, there is no evidence for shock in the Chabanais granite, which lies ~9 km from the crater's center, like there is at the Champagnac quarry, which lies ~6 km from the crater's center and has suevites directly covering the basement rocks. Finally, the serpentinite pole from the Merlis quarry (Site 4) lies near the 240 Ma pole on the Eurasian APWP [*Van der Voo*, 1993; *Torsvik et al.*, 2001], far from the population of suevites. Magnetic remanences from this site (>10 km from the crater's center) are most likely pre-impact.

A final point to address is whether tilting during post-impact rebound of the Rochechouart crater reoriented the magnetic directions of the suevites and basement rocks. Observations of most complex craters show that pre-impact rocks dip away from the crater's center with strikes that are tangential to the radius [*Grieve and Pilkington*, 1996; *Mazur et al.*, 2000; *Grieve and Therriault*, 2004]. Under this scenario, localities with southwest and up directions lying south of the crater that are subsequently tilted should have lower than average inclinations, while those to the north should have higher than average inclinations. This is not observed in the data (Table 1), which leads us to conclude that the magnetite remanences in the suevites were not significantly reoriented by post-impact rebound. The region has been tectonically quiescent since the impact.

The discovery of normal polarities in the Rochechouart suevites leads us to re-examine the multiple impact hypothesis of *Spray et al.* [1998]. If the suevites sampled in this study indeed averaged secular variation and have not been reoriented since cooling, we can then compare the overall mean pole from Rochechouart with that from Manicouagan in a Late Triassic reference frame. Two independent studies reported paleomagnetic data from Manicouagan. *Larochelle and Currie* [1967] published five site-mean VGPs from 44 cores based on a blanket demagnetization step at 20 mT while *Robertson* [1967] published six site-mean VGPs from 33 cores based on a blanket demagnetization step at 35 mT. We averaged the 11 site-means to calculate an overall mean pole of $\lambda= 58.9°N$, $\phi= 90.3°E$ ($A_{95}= 5.8°$) for the Manicouagan crater. Figure 3a shows a paleogeographic reconstruction of Eurasia and North America at 210 Ma using the parameters from *Van der Voo* [1993] and *Besse and Courtillot* [2002]. In this reference frame, the Rochechouart and Manicouagan poles are indistinguishable at 95% confidence limits. Figure 3b shows an Euler rotation at 210 Ma of Eurasia relative to a fixed North America following *Torsvik et al.* [2001]. In this reference frame, the Rochechouart and Manicouagan poles are again indistinguishable at 95% confidence limits. This finding, together with the dual magnetic polarities now recognized at Rochechouart, puts the bulk of the evidence in favor of the *Spray et al.* [1998] hypothesis for a coeval Late Triassic impact event on Earth.

**Acknowledgements**

We thank Rodger J. Hart for help in the field and Claude Marchat from the Pierre de Lune Association for his guiding us around the Rochechouart crater. Suggestions by J. Besse, Y. Gallet, S. T. Stewart and an anonymous reviewer improved the manuscript. IPGP publication #2162.

| Site | Name | Lithology | n/N/C | Dec | Inc | k | $\alpha_{95}$ | $P_{lat}$ | $P_{long}$ | dp | dm | $S_{lat}$ | $S_{long}$ |
|---|---|---|---|---|---|---|---|---|---|---|---|---|---|
| 1&2 | Champagnac quarry | Gran., leuco. & pseudo. | 7/9/8 | 230.3 | -20.8 | 63.6 | 7.6 | 34.8 | 113.7 | 3.3 | 6.2 | 45.855 | 0.833 |
| 3 | Champonger quarry | Metamorphic & granite | 0/7/7 | | | | | | | | | 45.835 | 0.771 |
| 4 | Merlis quarry | Serpentinite | 5/5/4 | 203.5 | -18.5 | 243.0 | 4.9 | 48.5 | 144.4 | 2.1 | 4.0 | 45.750 | 0.833 |
| 5 | Montoume quarry | Suevite | 12/12/10 | 219.2 | -43.5 | 113.7 | 4.1 | 52.7 | 110.4 | 2.5 | 4.0 | 45.776 | 0.775 |
| 6 | Fonceverane forest | Suevite | 6/6/4 | 33.1 | 40.8 | 174.4 | 5.1 | | | | | 45.812 | 0.769 |
| 7 | Fonceverane forest | Suevite | 3/5/3 | 29.0 | 46.4 | 44.5 | 18.7 | | | | | 45.812 | 0.769 |
| 6&7 | Fonceverane forest | Suevite | 9/11/7 | 31.8 | 42.7 | 93.3 | 6.9 | 57.0 | 119.5 | 4.1 | 6.6 | 45.812 | 0.769 |
| 8 | Fonceverane | Suevite | 4/4/3 | 215.9 | -41.8 | 177.8 | 6.9 | 53.9 | 115.4 | 5.2 | 8.5 | 45.815 | 0.763 |
| 9 | Fonceverane | Metamorphic | 0/3/3 | | | | | | | | | 45.815 | 0.763 |
| 10 | D160 road | Granite | 3/3/3 | 345.8 | 62.1 | 18.4 | 29.6 | | | | | 45.848 | 0.734 |
| 11 | Chabanais quarry | Granite | 4/4/3 | 228.0 | -6.8 | 145.8 | 7.6 | 30.6 | 121.3 | 2.9 | 5.8 | 45.879 | 0.729 |
| 12 | N141 highway | Granite | 0/3/3 | | | | | | | | | 45.909 | 0.904 |

**Table 1.** Paleomagnetic results for the Rochechouart impact melt breccias and basement rocks.

Abbreviations/notations are: Gran., granodiorite; leuco., leucogranite; pseudo., pseudotachylite; n, number of samples considered in the site mean direction; N, number of samples demagnetized; C, number of cores sampled; Dec - Inc, declination - inclination of the site-mean direction; k, the best estimate of the precision parameter, which is a measure of the dispersion of a population of directions; $\alpha_{95}$, the radius that the mean direction lies within 95% confidence; $P_{lat}$ - $P_{long}$, latitude - longitude of the virtual geomagnetic pole (VGP) in °N and °E; dp - dm, the minor and major axes of the ellipsoid containing the mean direction within 95% confidence limits; $S_{lat}$ - $S_{long}$, latitude - longitude of the sampling site in °N and °E.

**Figure 1. a)** Simplified geologic map of the Rochechouart crater, France; the dashed circle represents the approximate limit of the crater [after *Chèvremont et al.*, 1996]. **b-i)** Zijderveld diagrams and normalized moments of typical impact melt breccias and basement rocks from the Rochechouart meteorite crater.

**Figure 2. a)** Stereonet plot of the individual sample and site-mean (stars with $\alpha_{95}$ confidence ellipses) directions. The grey star represents the international geomagnetic reference field and the grey diamond represents the present-day geocentric axial dipole direction at Rochechouart. **b)** Polar projection of the site-mean virtual geomagnetic poles (VGPs) from Rochechouart (locality represented by dark star with R) and Manicouagan (locality represented by grey star with M and VGP by a grey plus; data from *Larochelle*

*and Currie* [1967] and *Robertson* [1967]). All VGPs are projected in the northern hemisphere. In light grey is the Eurasian apparent polar wander path (APWP) from *Van der Voo* [1993] and *Besse and Courtillot* [2002], in dark grey is the Eurasian APWP from *Torsvik et al.* [2001]—numbers next to the poles are time-window ages in millions of years. The four VGPs in light grey are from *Pohl and Soffel* [1971]. The melting degree of the suevites increases from the Chassenon-type (square), to the Montoume-type (triangle) and finally to the Babaudus-type (circle) [*Lambert*, 1974; 1977b; *Chèvremont et al.*, 1996]. The trajectory of these VGPs possibly record geomagnetic secular variation ~214 million years ago during progressive cooling of the impact melt breccias.

**Figure 3. a)** Paleogeographic reconstruction at 210 Ma using the parameters from *Besse and Courtillot* [2002] and *Van der Voo* [1993]. The Eurasian APWP is in light grey and the North American APWP is in dark grey— numbers next to the poles are time-window ages in millions of years. The overall mean pole from Rochechouart (data from this study; locality represented by dark star with R) and Manicouagan (locality represented by grey star with M) are indistinguishable at 95% confidence limits (Manicouagan data from *Larochelle and Currie* [1967] and *Robertson* [1967]). **b)** Paleogeographic reconstruction at 210 Ma in a North American-fixed reference frame using the parameters from *Torsvik et al.* [2001]. In grey is the combined North American and European APWP [*Torsvik et al.*, 2001].

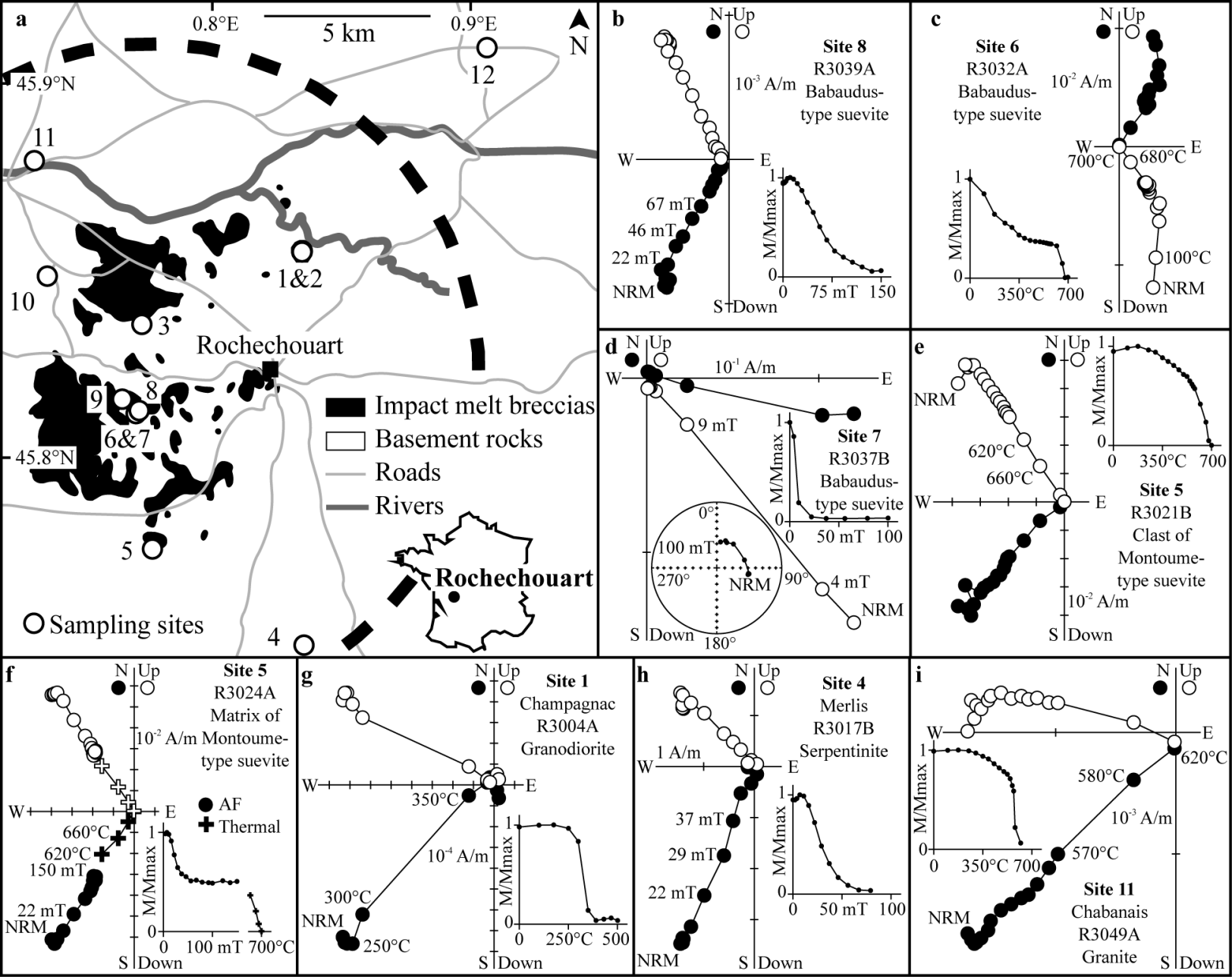

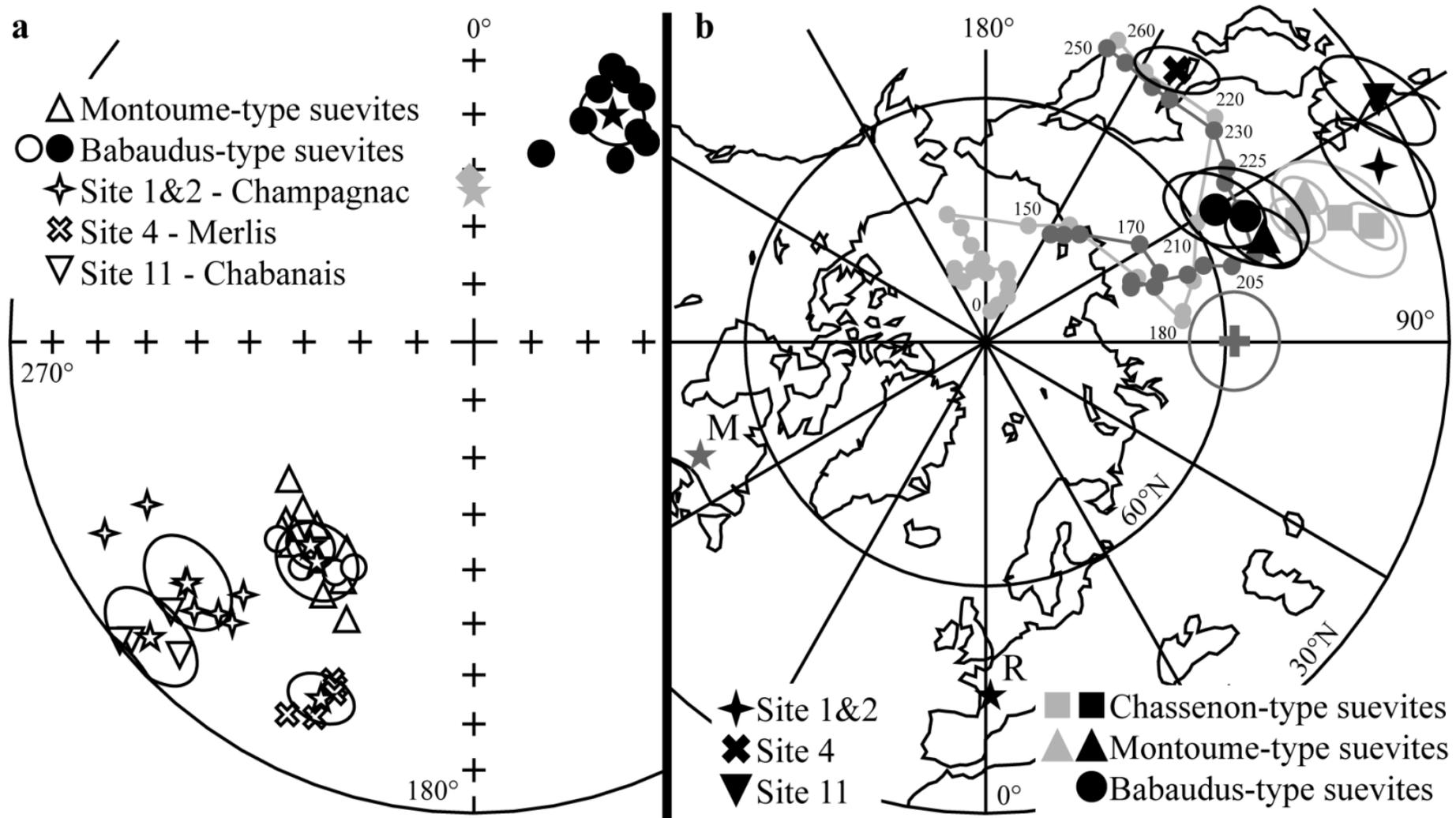

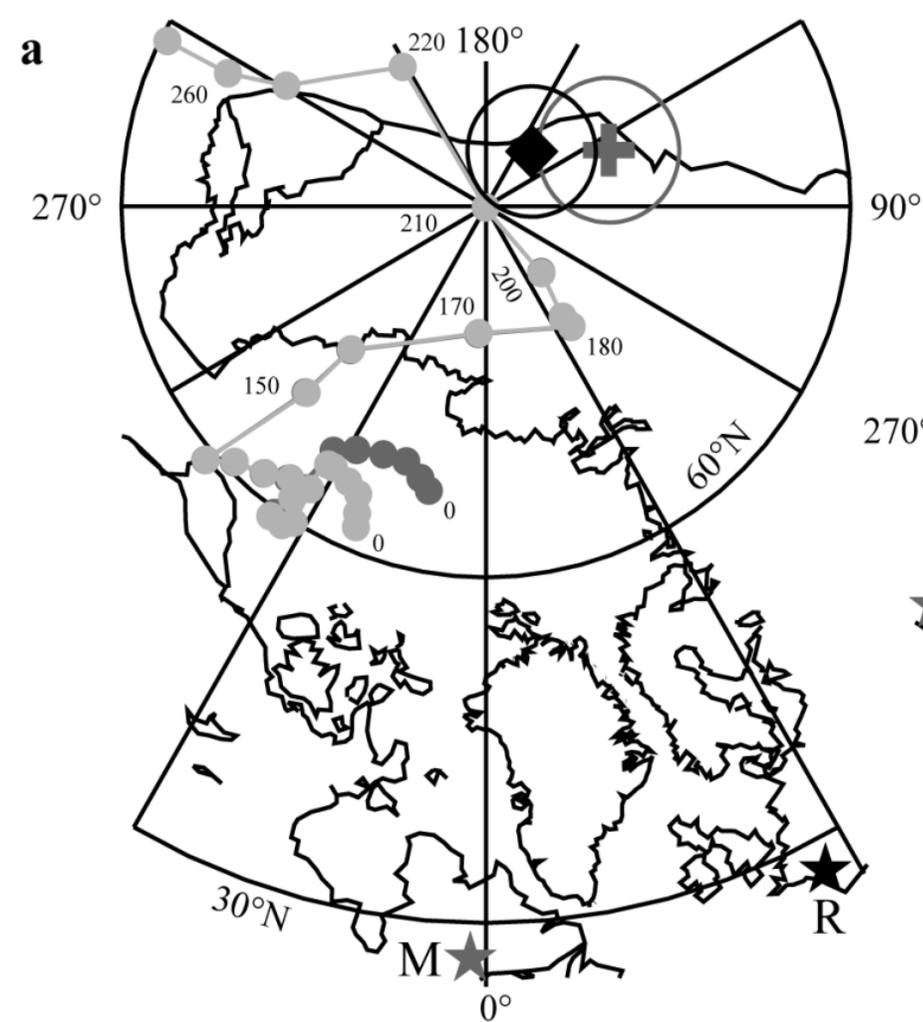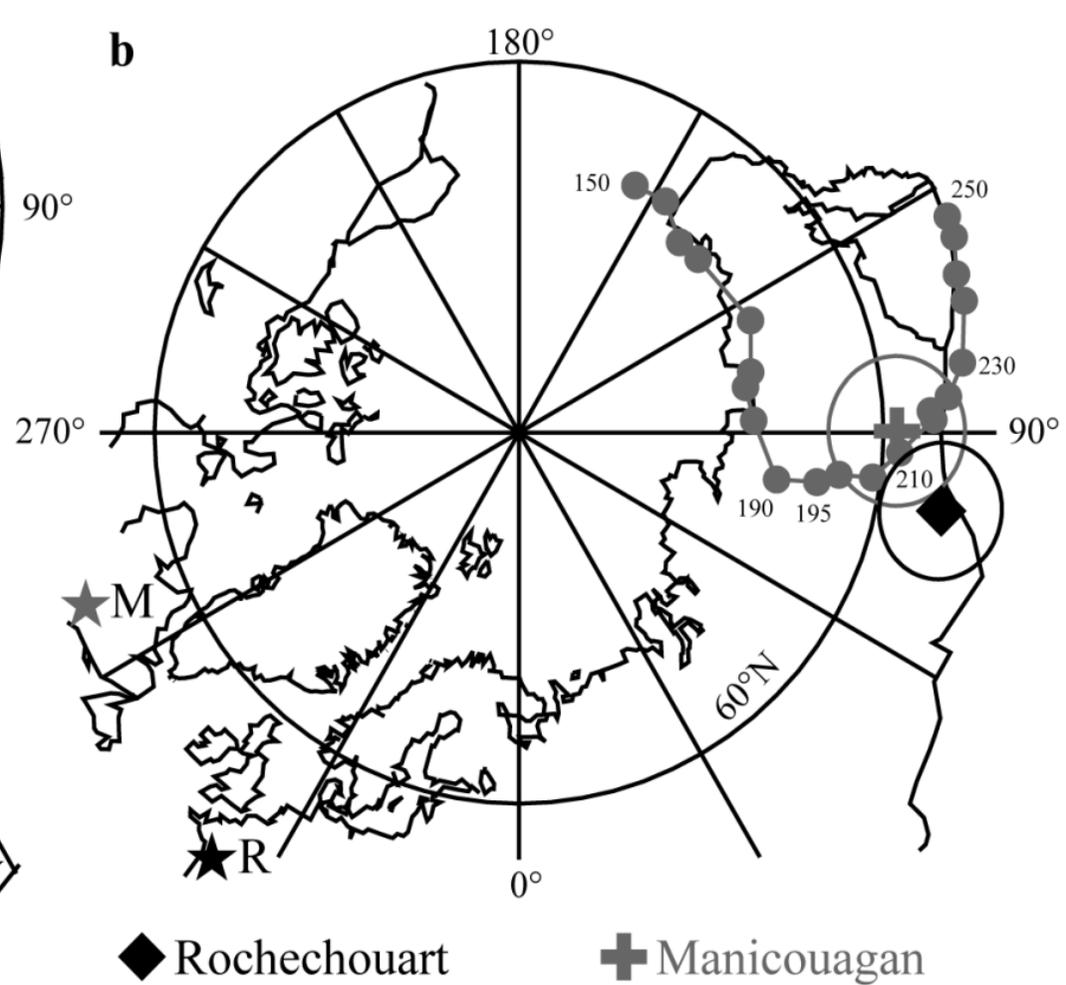